\documentclass[useAMS,usenatbib,dvipdfmx]{mn2e}
\usepackage{setspace, ulem}
\usepackage[dvipdfmx]{graphicx,color}
\title[Transonic outflows in a dark matter halo with a central black hole]
{Transonic galactic outflows in a dark matter halo with a central black hole and its application to the Sombrero galaxy}
\author[Igarashi, Mori and Nitta]{Asuka Igarashi$^{1}$, Masao Mori$^{1}$ and Shin-ya Nitta$^{2,3,4}$\\
$^{1}$University of Tsukuba, 1-1-1, Tennodai, Tsukuba, Ibaraki, 305-8577, Japan\\
$^{2}$Tsukuba University of Technology, 4-3-15, Amakubo, Tsukuba, Ibaraki, 305-8520, Japan\\
$^{3}$Hinode Science Project, National Astronomical Observatory of Japan, 2-21-1 Osawa, Mitaka, Tokyo, 181-8588, Japan\\
$^{4}$Institute of Space and Astronautical Science, Japan Aerospace Exploration Agency, 3-1-1 Yoshinodai, Sagamihara, \\Kanagawa 229-8510, Japan}
\begin{document}

\date{Accepted 201x ***. Received 2014 ***; in original form 2014 ***}

\pagerange{\pageref{firstpage}--\pageref{lastpage}} \pubyear{201x}

\maketitle

\label{firstpage}

\begin{abstract}

We have classified possible transonic solutions of galactic outflows in the gravitational potential of the dark matter halo (DMH) and super massive black hole (SMBH) under the assumptions of isothermal, spherically symmetric and steady state. 
It is clarified that the gravity of SMBH adds a new branch of transonic solutions with the transonic point in very close proximity to the centre in addition to the outer transonic point generated by the gravity of DMH. 
Because these two transonic solutions have substantially different mass fluxes and starting points, these solutions may have different influences on the evolution of galaxies and the release of metals into intergalactic space. 
We have applied our model to the Sombrero galaxy and obtained a new type of galactic outflow: a slowly accelerated transonic outflow through the transonic point at very distant region ($\simeq 126$\ kpc). 
In this galaxy, previous works reported that although the trace of the galactic outflow is observed by X-ray, the gas density distribution is consistent with the hydrostatic state.
We have clarified that the slowly accelerating outflow has a gas density profile quite similar to that of the hydrostatic solution in the widely spread subsonic region. Thus, the slowly accelerating transonic solution cannot be distinguished from the hydrostatic solution in the observed region ($\leq 25$\ kpc) even if slow transonic flow exists. 
Our model provides a new perspective of galactic outflows and is applicable even to quiescent galaxies with inactive star formation.

\end{abstract}


\section{Introduction}

Recent observational cosmology reveals that the primary component of galactic mass is cold dark matter (CDM), which plays essential roles in the evolution of galaxies and the large-scale structure of the universe. 
So far, studies of galaxy formations based on the CDM scenario indicate that galactic outflows play significant roles in the evolution of galaxies and  metal enrichment of intergalactic space. 

It is well known that the ratio of gas mass to stellar mass in elliptical galaxies is smaller than that in spiral galaxies \citep{osterbrock60}.
This deficiency of gas in elliptical galaxies indicates that galactic outflows efficiently remove interstellar gas from these galaxies \citep{johnson71,mathews71}.
In addition, the low-density intergalactic medium at high-$z$ contains a small amount of metals \citep{songaila97,aguirre01,ellison00}.
The existence of metals in intergalactic space suggests that galactic outflows transport metals to this space, because the metals should come from stars inside galaxies.
These observed results strongly indicate the importance of galactic outflows in the evolution of galaxies. 
To realize galactic outflows, a sufficient thermal energy supply is required to escape from the gravitational potential well of the galaxy \citep{larson74,dekel86,mori97,mori99,mori02,binney04,cattaneo06,oppenheimer06,puchwein12}.
Thus far, studies have mainly assumed supernovae and stellar winds as the thermal energy sources.
Active galactic nuclei (AGN) and cosmic rays have also been presented as sources of thermal energy for driving galactic outflows \citep{silk98,sharma12,sharma13}.
Thus, from the theoretical perspective, there are many possible sources of thermal energy to drive galactic outflows.

In the local universe, a significant number of star-forming galaxies form starburst-driven outflows \citep{strickland02,heckman03}.
Furthermore, high-redshift galaxies show that the outflow velocity is proportional to the star formation rate and the stellar mass of galaxies \citep{kennicutt98a,strickland04,pettini01,shapley03,weiner09}.

In contrast, we intend to focus on transonic solutions as galactic outflows without starbursts in this paper.
In our model, transonic galactic outflows are driven by energy stored in interstellar gas itself.
\citet{parker58} first examined spherically symmetric solar wind and clarified that this solar wind can become supersonic because they begin at the subsonic region and include the transonic point.
This result shows that the energy supply and the gravity are essential for forming the transonic solution.
This transonic solar wind has been actually observed in the outflow from the sun, as evidenced by the success of {\it Mariner $II$} in \citep{neugebauer62}, and it is well known as the entropy-maximum solution connecting the starting point (sun) to infinity \citep{lamers99}. 

Several studies have examined transonic galactic outflows in terms of solar wind. \citet{burke68} and \citet{johnson71} applied the solar wind model to galactic-scale transonic outflows with the stellar halo mass distribution.
\citet{wang95} studied the nature of galactic outflows with DMH mass distribution including radiative cooling.
However, owing to an unrealistic assumption of a single power-law DMH mass distribution ($\propto r^{-2}$), they could not realize transonic solutions.
Thus, determination of transonic solutions remains elusive.

\citet{sharma13} studied steady and spherically symmetric transonic galactic outflows in star-formation galaxies in the gravitational potential of a CDM halo not including stellar mass.
They assumed that the thermal energy was from supernovae and AGN radiation.
As a result, they concluded that AGN is important for driving high-velocity outflows and that galactic outflows control the observed correlation between DMH mass and stellar halo mass.
However, they assumed fixed transonic points for transonic outflows, although the transonic points should be determined naturally by the ratio of the thermal energy supply to the gravitational potential. 

\citet{tsuchiya13} discussed the influence of gravity from DMH mass distribution on the nature of the transonic solutions.
They assumed DMH mass for the gravitational potential because the dark matter is the dominant component of the gravity source in galaxies.
However, the functional form of DMH mass distribution has not reached a consensus, and many simulations have suggested different mass profiles.
For example, on the basis of the CDM scenario, \citet{navarro96} concluded by using the Navarro-Frenk-White (NFW) model that the structure of the DMH distribution is expressed by a double power law, $\rho_\mathrm{DMH}\propto r^{-1}(r+r_\mathrm{d})^{-2}$, where $r$ is radius from the centre of the galaxy and $r_\mathrm{d}$ is the scale length of the DMH.
Other simulations with higher resolution also supported the double power-law mass distribution in the CDM scenario, although the power-law index was somewhat different.
For example, \citet{fukushige97} and \citet{moore99} suggested $\rho_\mathrm{DMH}\propto r^{-1.5}(r^{1.5}+r_\mathrm{d}^{1.5})^{-1}$.
These numerical models commonly have a ``cusp" of mass distribution at the centre.
In contrast, observations of nearby dwarf galaxies have suggested that their DMH distributions have a flat ``core" at the centre. \citet{burkert95} proposed an empirical profile with the core structure as $\rho_\mathrm{DMH}\propto (r+r_\mathrm{d})^{-1}(r^2+r_\mathrm{d}^2)^{-1}$.
This unsolved problem is the so-called core-cusp problem \citep{moore99}. 

\citet{tsuchiya13} adopted a functional form of the mass distribution as $\rho_\mathrm{DMH}\propto r^{-\alpha}(r+r_\mathrm{d})^{-3+\alpha}$ with a concentration parameter $\alpha$ to denote the variety of the distribution at the centre.
This profile approximately reproduces the Burkert profile with $\alpha=0$, the NFW profile with $\alpha=1$ and the Moore profile with $\alpha=1.5$, as discussed in Section 2.1.
By using this distribution model of DMH, \citet{tsuchiya13} first reported transonic solutions with the gravitational potential of DMH. 
In addition, they showed the possibility of a new type of transonic solution in which the transonic point forms in a very distant region ($\sim100$\ kpc).

However, it is widely accepted that most galaxies have a central super massive black hole (SMBH) \citep{marconi03}. The gravitational potential of SMBH must influence the nature of galactic outflows at the central region. Therefore, in this paper, we add the gravity of the central SMBH contribution to the Tsuchiya model and analyse the transonic solutions, whereas the original Tsuchiya model includes only the gravity of DMH. To  construct our model, we assume an isothermal, spherically symmetric and steady state without the injection of mass along the outflow lines, as reported by \citet{tsuchiya13}, with the aim of clarifying transonic solutions under realistic mass distribution models including both DMH and SMBH. 

In addition, we apply our model to the Sombrero galaxy (NGC4594) to determine the acceleration process of the galactic outflows in an actual galaxy. \citet{li11} reported a contradiction in that although the trace of the galactic outflow is observed by X-ray, the gas density distribution in this galaxy is well reproduced as the hydrostatic state. 
They argued a possibility that the hot gas forms a subsonic outflow, the thermal structure of which resembles, and is difficult to distinguish from a quasi-hydrostatic state. We attempt to resolve this discrepancy with our model. 

The structure of this paper is as follows. In Section 2, we summarize the results of \citet{tsuchiya13} in Section \ref{result of tsuchiya} and the influence of the central black hole in Section \ref{result  of model}. In Section 3, we discuss the application to the Sombrero galaxy. In Section 4, we discuss several topics, and we give the conclusion in Section 5. 

\begin{figure*}
 \centering
 \includegraphics[width=1.5\columnwidth]{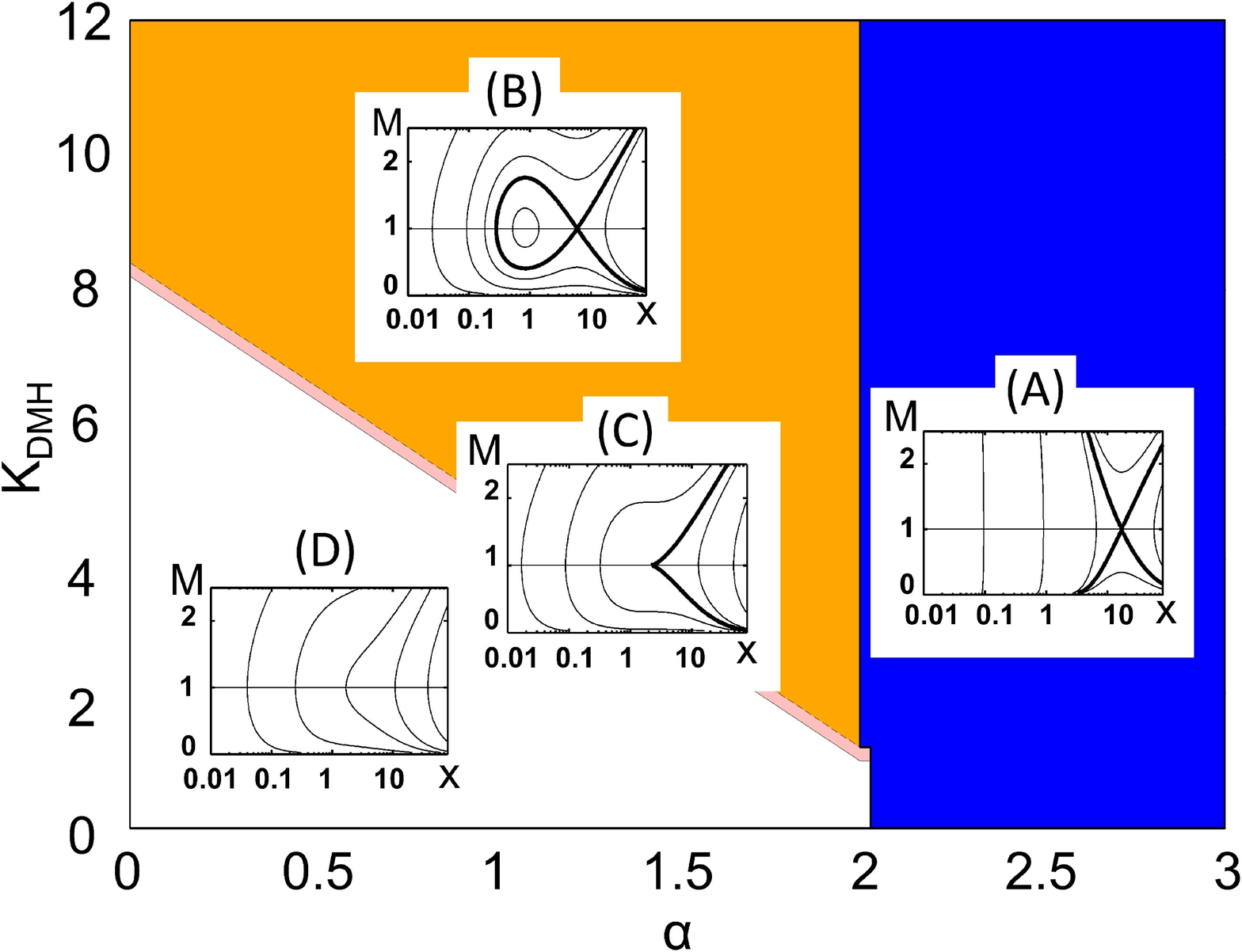}
 \caption{Various solutions of the model with dark matter halo (DMH) distributions \citep{tsuchiya13}. The horizontal axis shows the power-law index of DMH, and the vertical axis is $K_\mathrm{DMH}$ defined in Eq. (\ref{kandkbh1}). The inset panels (A), (B), (C) and (D) represent transonic solutions (thick bold lines) and other solutions (thin lines). The panel (A) in the blue region has only one X-point, (B) in the orange region has a single X-point with a single O-point, (D) in the white region has no critical point. The panel (C) on the magenta line is the border solution. }
 \label{fig-tsuchiya}
\end{figure*}

\section[]{Analytical Model for Transonic Outflow}

In this section, we first summarize results of \citet{tsuchiya13}, then we show our model. Here, we assume an isothermal, steady, spherically symmetric outflow model, and we ignore mass injection along the flow except the starting point. 

\subsection{The model for this study}
The basic equations are those for the conservation of mass and momentum:
\begin{eqnarray}
&{}&4\pi\rho vr^2=\dot{M},\\
&{}&v\frac{\partial v}{\partial r}=-\frac{c_\mathrm{s}^2}{\rho}\frac{\partial \rho}{\partial r}-\frac{\partial \Phi}{\partial r},
\end{eqnarray}
where $\rho$, $v$, $r$, $\dot{M}$, $c_\mathrm{s}$ and $\Phi$ are gas density, gas velocity, radius from the centre, mass flux, sound speed and gravitational potential, respectively. $\dot{M}$ is a constant because we ignore the mass injection along the flow except at the starting point. $c_\mathrm{s}$ is also a constant because of the isothermal approximation. Substituting $\rho$ from Eqs. (1) and (2), we obtain
\begin{eqnarray}
&{}&\frac{d \mathcal{M} ^2}{d x}=\frac{N(x)}{1-\frac{1}{\mathcal{M}^2}},\label{tsuchiya-eq}\\
&{}&N(x)=\frac{4}{x}-\frac{2}{c_\mathrm{s}^2}\frac{d\Phi}{dx},\label{tsuchiya-n}
\end{eqnarray}
where $\mathcal{M}$ is the Mach number, which is equal to $v/c_\mathrm{s}$; $x$  is the non-dimensional radius defined as $x=r/r_\mathrm{d}$; and $r_\mathrm{d}$ is the scale length of DMH mass distribution.

\citet{tsuchiya13} adopted a model of the mass density profile of DMH as 
\begin{eqnarray}
\rho_\mathrm{DMH}(r;\alpha,r_\mathrm{d},\rho_\mathrm{d})=\frac{\rho_\mathrm{d}r_\mathrm{d}^3}{r^{\alpha}(r+r_\mathrm{d})^{3-\alpha}}, \label{dmh model}
\end{eqnarray}
where $\rho_\mathrm{d}$ represents the scale density. They assumed a simple gravitational potential model that includes the contribution only from DMH, as defined as Eq. (\ref{dmh model}). In the limit of $x\rightarrow 0$, $\rho_\mathrm{DMH} \propto r^{-\alpha}$ (0$<\alpha<$3) and $\rho_\mathrm{DMH} \propto x^{-3}$ for $x\rightarrow \infty$. This model well and approximately reproduces various models predicted from both theoretical and observational perspectives. For example, this model corresponds exactly to the NFW model \citep{navarro96} with $\alpha=1$ and approximately to the Moore model \citep{moore99,fukushige97} with $\alpha=1.5$ and the Burkert model \citep{burkert95} with $\alpha=0$. The plausible value of the index $\alpha$ remains an open question. Thus, we treat the concentration parameter $\alpha$ as a flexible parameter in this study. 

We add the gravitational potential of central SMBH to the potential model in \citet{tsuchiya13}. Eq. (\ref{tsuchiya-n}) becomes 
\begin{equation}
N(x)=\frac{4}{x}-\frac{4K_\mathrm{DMH}}{x^2}\int_\mathrm{0}^x x^{2-\alpha} (x+1)^{\alpha-3}dx+\frac{4K_\mathrm{BH}}{x^2}, 
\end{equation}
where 
\begin{eqnarray}
&{}&K_\mathrm{DMH}=\frac{2\pi G\rho_\mathrm{d}r_\mathrm{d}^2}{c_\mathrm{s}^2},\label{kandkbh1}\\
&{}&\hspace{2.5mm}K_\mathrm{BH} =\frac{GM_\mathrm{BH}}{2r_\mathrm{d}c_\mathrm{s}^2}, \label{kandkbh2}
\end{eqnarray}
 $G$ and $M_\mathrm{BH}$ are the Newton constant and the mass of SMBH, respectively. Integrating Eq. (\ref{tsuchiya-eq}), we obtain
\begin{eqnarray}
\mathcal{M}^2-\log \mathcal{M}^2=4\log x-4\Phi'(x;\alpha,K_\mathrm{DMH},K_\mathrm{BH})+C, \label{thisstudy-mach}
\end{eqnarray}
and
\begin{eqnarray}
&{}&\Phi'(x;\alpha,K_\mathrm{DMH},K_\mathrm{BH}) \nonumber\\
&{}&=\frac{1}{2c_\mathrm{s}^2}\Phi \nonumber\\
&{}&=K_\mathrm{DMH}\int \frac{1}{x^2}\left(\int_\mathrm{0}^x x^{2-\alpha}(x+1)^{\alpha-3}dx\right)dx -\frac{K_\mathrm{BH}}{x}, \nonumber\\
&{}& \label{Phi'}
\end{eqnarray}
where $C$ is the integration constant.

\begin{figure*}
 \centering
 \includegraphics[width=2.\columnwidth]{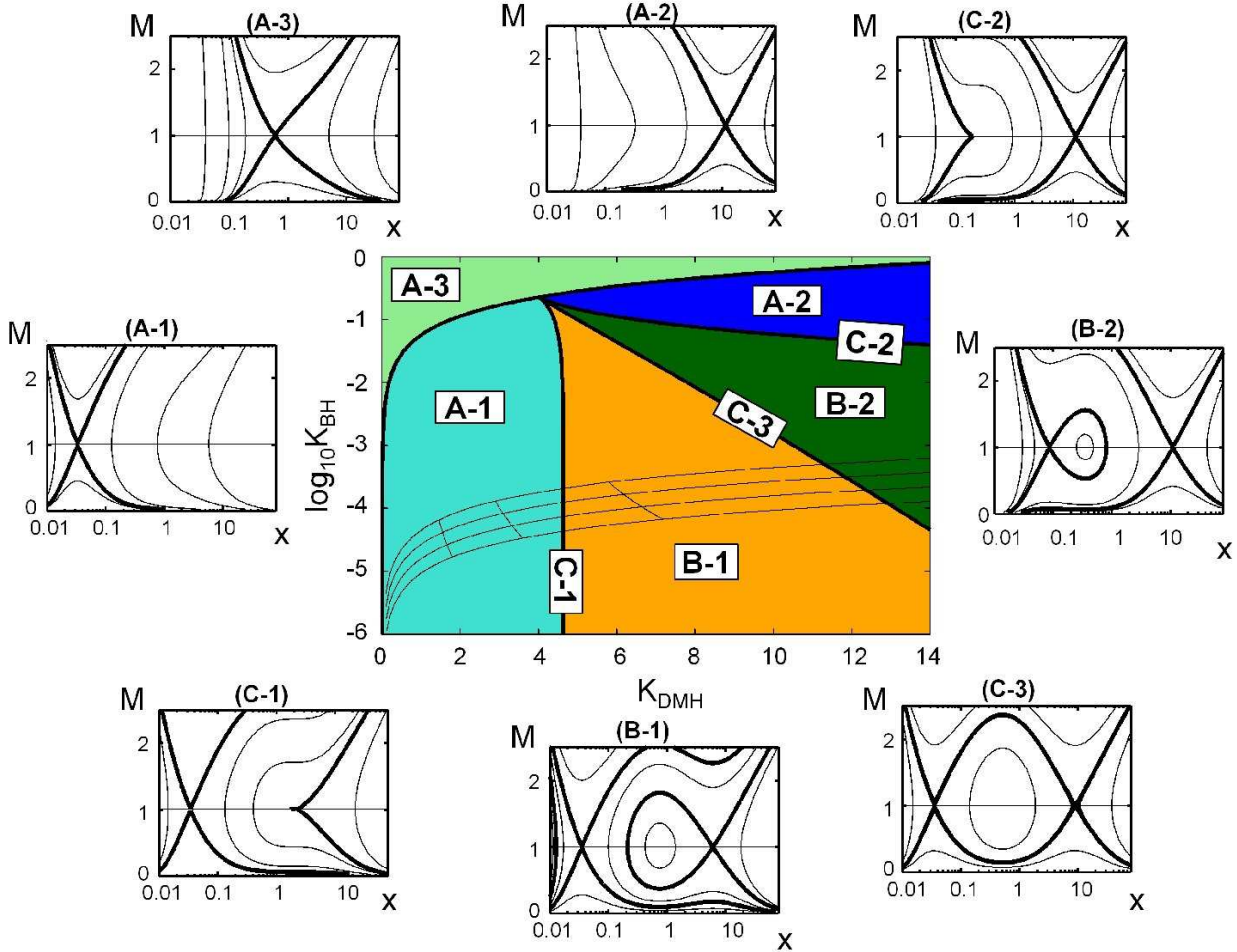}
  \caption{Various solutions of the model with dark matter halo (DMH) ($\alpha=1$) and the central black hole. The horizontal axis is $K_\mathrm{DMH}$ defined in Eq. (\ref{kandkbh1}) and roughly corresponds to the ratio of the gravitational potential of DMH and the thermal energy. The vertical axis is $K_\mathrm{BH}$ defined in Eq. (\ref{kandkbh2}) and corresponds to the ratio of the gravitational potential of SMBH and the thermal energy. The labels, such as A-1, represent the types of transonic solutions presented in Table \ref{table-thisstudy}. The hatched region represents the range of actual galaxies (defined in Eqs. (16) and (17) with $(\kappa,\xi,\mu,\nu)=(9.7,-0.074,0.11,1.27)$; see Section 4.2.1.). The four thin solid lines represent the range of actual galaxies from $10^{11}\ \mathrm{M}_\mathrm{\odot}$(bottom) to $10^{14}\ \mathrm{M}_\mathrm{\odot}$(top). The three thin solid lines intersecting the four solid lines represent $\eta=0.5$, $1$ and $2$ from left. \citep{prada12,baes03} }
 \label{fig-thisstudy10}
\end{figure*}

\subsection{Transonic solutions in the gravitational potential of DMH without SMBH}\label{result of tsuchiya}
\citet{tsuchiya13} considered only the effect of the gravity of DMH, which corresponds to $K_\mathrm{BH}=0$ in Eq. (\ref{thisstudy-mach}). They determined two types of transonic solutions having single X-point (transonic point) with single O-point and that with only one X-point (Figs. 2. and 4. of Tsuchiya et al. 2013). We summarize these solutions in Fig. \ref{fig-tsuchiya} by virtue of two parameters, $K_\mathrm{DMH}$ and $K_\mathrm{BH}$. The horizontal axis shows the power index $\alpha$, and the vertical axis shows $K_\mathrm{DMH}$ defined in Eq. (\ref{kandkbh1}). $K_\mathrm{DMH}$ corresponds approximately to the ratio of the gravitational potential energy density ($2\pi G \rho_\mathrm{d} r_\mathrm{d}^2$) to the thermal energy density ($c_\mathrm{s}^2$). 

When $\alpha <2$ with small $K_\mathrm{DMH}$, there is no transonic solution (white region in Fig.\ref{fig-tsuchiya}). When $\alpha >2$, there is a transonic solution with a single X-point (blue region in Fig. \ref{fig-tsuchiya}). This transonic solution begins at the centre and extends to infinity. When $\alpha <2$ with large $K_\mathrm{DMH}$, there is also a transonic solution with an X-point and an O-point (orange region in Fig. \ref{fig-tsuchiya}). This transonic solution also extends to infinity but does not begin at the centre. When the concentration parameter of DMH $\alpha$ increases, the position of the X-point moves outward and that of the O-point moves inward. When $c_\mathrm{s}^2$ corresponding approximately to the thermal energy decreases or $2\pi G \rho_\mathrm{d} r_\mathrm{d}^2$ corresponding approximately to the gravitational potential energy increases, the position of the X-point moves to the outward region and that of the O-point moves inward.

\subsection{The effect of SMBH to transonic solutions}\label{result of model}

In this section, we estimate the effect of the gravitational potential of SMBH to the transonic solutions. The mass density distribution of DMH is widely spread, and the gravitational potential of DMH is dominant in a huge scale of a galaxy. On the contrary, the gravitational potential of SMBH is dominant in the vicinity of the centre. Because of these properties of gravitational potentials, the mass of SMBH must be added to our galactic outflow model to construct a realistic gravitational field model near the centre. 

We summarized the obtained transonic solutions for $K_\mathrm{BH} \neq 0$ in phase diagrams. Fig. \ref{fig-thisstudy10} shows the phase diagram for $\alpha=1$; phase diagrams for other values of $\alpha$ are shown in Fig. 3. Two types of transonic solutions are apparent, including that with a single X-point and that with a pair of two X-points with one O-point (Fig. \ref{fig-thisstudy10}). We define the former type as A and the latter type as B. Type A has one transonic solution such as the Parker solution \citep{parker58}. On the contrary, type B has two transonic solutions. The solutions on the boundaries in Fig. 3 are defined as type C. The details of the aforementioned solutions are summarized in Table \ref{table-thisstudy}. 

If the locus of the extreme points of $N(x)$ from the centre is longer/shorter than that of the single X-point, the X-point is formed mainly by the gravity of  SMBH/DMH. This transonic solution formed by SMBH/DMH is shown in the region A-1/A-2 in Fig. \ref{fig-thisstudy10}. The region with no extreme points is shown as region A-3 in the figure. These transonic solutions begin at the centre and extend to infinity.

When there are two X-points with a single O-point (type B), the inner/outer X-point is formed by SMBH/DMH. The transonic solution through the inner/outer X-point is referred to as type $\rmn{X_{in}/X_{out}}$ in this paper. Because $K_\mathrm{DMH}$ corresponds to the gravitational potential of DMH, the more $K_\mathrm{DMH}$ increases, the more the outer X-point moves outward. On the contrary, because $K_\mathrm{BH}$ corresponds to that of DMH, the more $K_\mathrm{BH}$ increases, the more the inner X-point moves outward. On the boundary between B-1 and B-2, a special transonic solution connecting two X-points appears (region C-3 in Fig. \ref{fig-thisstudy10}). In region B-1, type $\rmn{X_{in}}$ solution begins at the centre, but type $\rmn{X_{out}}$ solution does not. In region B-2, type $\rmn{X_{out}}$ solution extends to infinity, but type $\rmn{X_{in}}$ solution does not.

As $\alpha$ increases, the influence of DMH becomes more dominant in the central region. Thus, if $\alpha$ is large, the gravitational potential of DMH is more dominant than that of SMBH in the vicinity of the centre. We have determined that when $\alpha$ increases, the area of region A-2 extends and that of A-1 shrinks (Fig. 3). When $\alpha>2$, only type A-3 solution is present. 

\begin{figure}
 \centering
 \includegraphics[width=0.99\columnwidth]{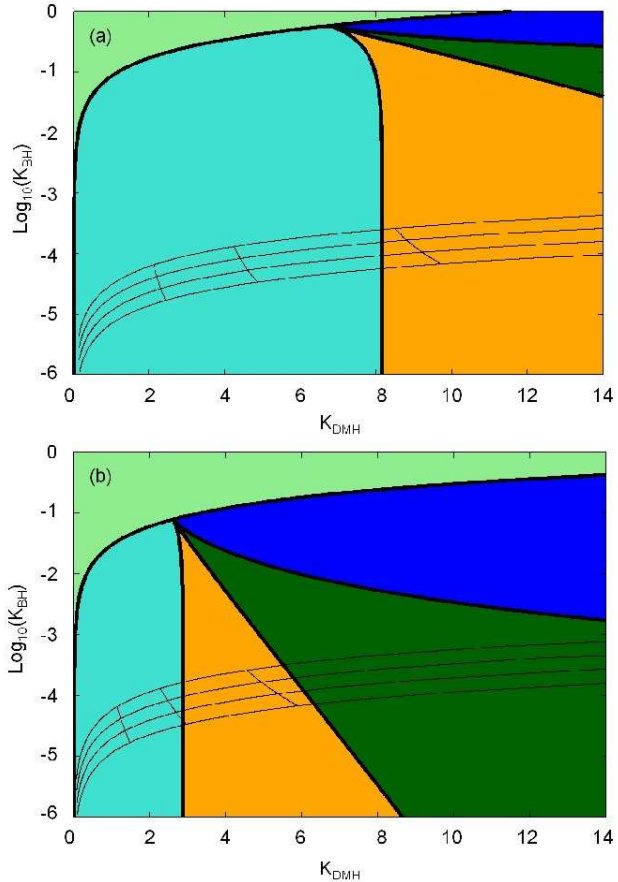}
  \caption{(a) Solution map of the model with dark matter halo (DMH) ($\alpha=0$) and central black hole. (b) Solution map of the model with dark matter halo ($\alpha=1.5$) and central black hole. When $\alpha$ increases, the influence of DMH becomes larger at the central region.}
 \label{fig-thisstudy00and15}
\end{figure}

\begin{table}
 \centering
 \begin{tabular}{c|c|c|c}\cline{1-3}
A & 1 & \parbox{7cm}{Single X-point generated by the central super massive black hole (SMBH).} & \ \\ \cline{2-3}
\ & 2 & \parbox{7cm}{Single X-point generated by the dark matter halo (DMH).} & \ \\ \cline{2-3}
\ & 3 & \parbox{7cm}{Single X-point with no extreme points in N(x).} & \ \\ \cline{1-3}
B & 1 & \parbox{7cm}{Two X-points with one O-point.} & \  \\
\ &\  & \parbox{7cm}{The transonic solution, through X-point generated by the DMH, rounds O-point and breaks off.} &  \\ \cline{2-3}
\ & 2 & \parbox{7cm}{Two X-points with one O-point.} \\
\ &\  & \parbox{7cm}{The transonic solution, through X-point generated by the SMBH, rounds O-point and breaks off.} & \ \\ \cline{1-3}
C & 1 & \parbox{7cm}{Boundary solution between A-1 and B-1.} &\  \\ \cline{2-3}
\ & 2 & \parbox{7cm}{Boundary solution between A-2 and B-2.} &\  \\ \cline{2-3}
\ & 3 & \parbox{7cm}{Boundary solution between B-1 and B-2.} &\  \\ \cline{1-3}
 \end{tabular}
 \caption{Features of solutions with the gravitational potential of dark matter halo (DMH) and central black hole.}
 \label{table-thisstudy}
\end{table}

\section{Application to the Sombrero Galaxy}\label{the sombrero galaxy}
In this section, we apply our model to the Sombrero galaxy (NGC4594) to determine the acceleration process of the galactic outflow in an actual galaxy.

The {\it Chandra X-ray Observatory} detected a diagnostic feature of the galactic outflow as diffuse hot gas in the Sombrero galaxy \citep{li11}. In fact, the estimated amount of the gas component in that galaxy ($M_\mathrm{HI}=3.18\times10^8\ \mathrm{M}_\mathrm{\odot}$, $M_\mathrm{H_2}=4.44\times10^8\ \mathrm{M}_\mathrm{\odot}$) is unnaturally small compared with the predicted gas mass supplied from stellar winds ($0.258-0.415\ \mathrm{M}_\mathrm{\odot}\mathrm{yr}^{-1}$) and type Ia supernovae ($0.1\ \mathrm{M}_\mathrm{\odot}\mathrm{yr}^{-1}$) \citep{bajaja84,bajaja91,athey02,knapp92,mannucci05,cappellaro99,li07,kennicutt98}. The star-forming rate, estimated as $0.06\ \mathrm{M}_\mathrm{\odot}\mathrm{yr}^{-1}$, is lower than that in other early type spiral galaxies \citep{li07,hameed05}. Thus, in comparison with the gas mass supply from which the star-forming rate is deducted, the mass of gas in this galaxy is deficient. This deficiency implies that the Sombrero galaxy experienced drastic extraction of gas owing to galactic outflow. 

On the contrary, the observed gas density distribution is similar to hydrostatic distribution \citep{li11}. This hydrostatic-like gas distribution may imply an absence of galactic outflow. Thus, this property is inconsistent with the aforementioned results in which the widely spread hot gas distribution and the deficiency of the gas mass imply the existence of galactic outflow \citep{li11}. We applied our model to this galaxy to resolve this discrepancy. 

For the application of our model to the Sombrero galaxy, the mass distribution of this galaxy was determined by observations (see Appendix \ref{The Mass Distribution of DMH in the Sombrero Galaxy}). The SMBH in the Sombrero galaxy is believed to be inactive \citep{heckman80}. Thus, the SMBH influences only to gravitational potential. The observed averaged gas temperature is approximately $0.6 \,\rmn{keV}$ up to $25$\ $\rmn{kpc}$ \citep{li11}.

\begin{figure*}
 \centering
 \includegraphics[width=2\columnwidth]{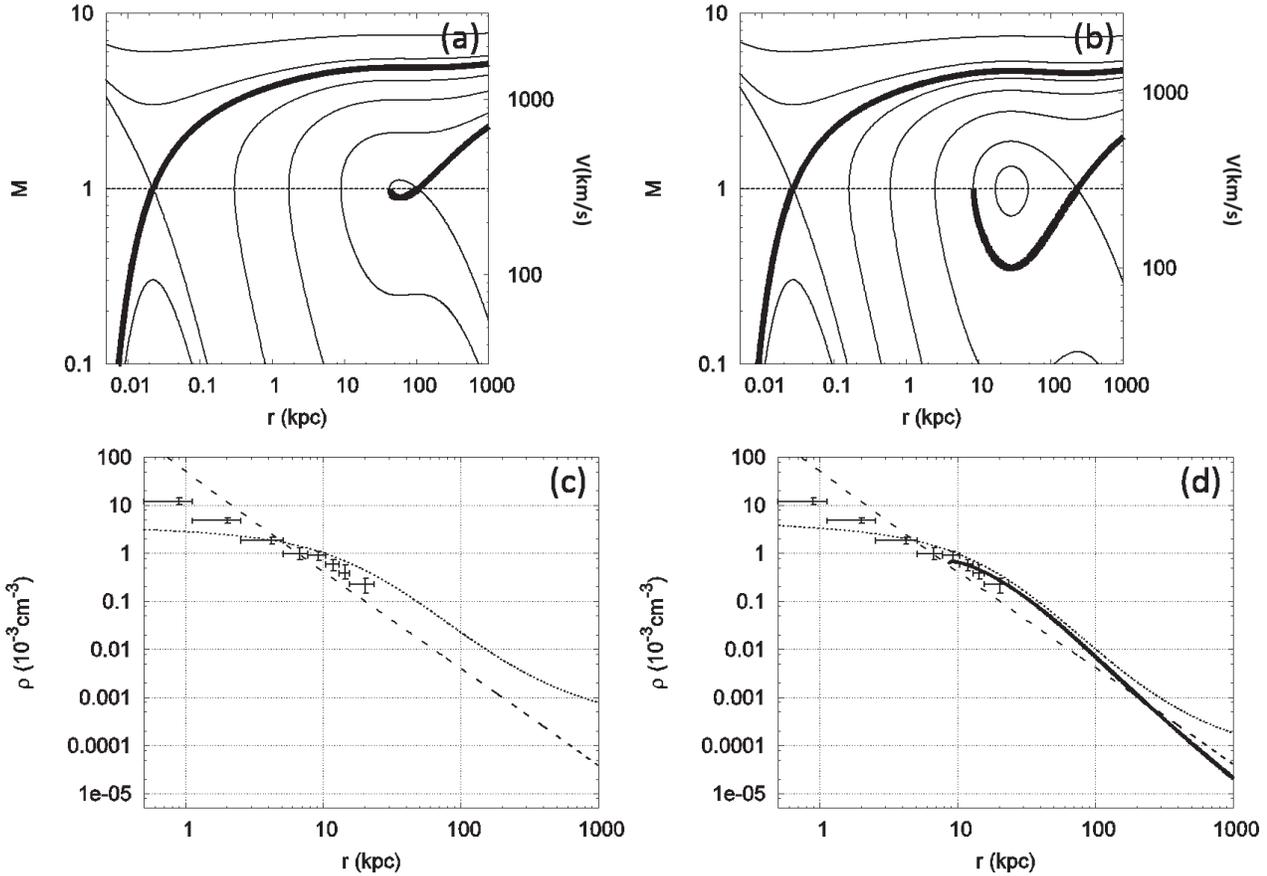}
  \caption{Mach number and gas density distribution in the gravitational potential of the dark matter halo (DMH) and super massive black hole (SMBH). We use parameters of the Sombrero galaxy: $\alpha=1,r_\mathrm{d}=36.1\ \mathrm{kpc}$, $\log_\mathrm{10}M_\mathrm{25}/\mathrm{M}_\mathrm{\odot}= 11.95$ and $M_\mathrm{BH}=10^9\ \mathrm{M}_\mathrm{\odot}$. (a) Mach number radial profiles of 0.6 keV. Bold solid lines represent transonic solutions. (b) Mach number radial profiles for 0.5 keV. (c) Radial density profile for 0.6 keV. Dotted curves represent the hydrostatic state. Dashed curves represent the transonic outflow that passes through the inner X-point. (d) Radial density profile for 0.5 keV. The solid curve represents the transonic outflow that passes through the outer X-point.}
 \label{fig-sombrero}
\end{figure*}

\begin{table}
\begin{center}
 \caption{Critical points in the Sombrero galaxy with the gravitational potential of the dark matter halo (DMH) and super massive black hole (SMBH).}\label{tb1}
  \begin{tabular}{|c|c|c|c|}
    &$r_\mathrm{BH}(\rmn{kpc})$&$r_\mathrm{O}(\rmn{kpc})$&$r_\mathrm{DMH}(\rmn{kpc})$ \\ \hline
   $0.6$ $\rmn{keV}$&0.021&58.0&105.4\\ \hline
   $0.5$ $\rmn{keV}$&0.025&28.0&232.2\\ \hline
  \end{tabular}
\end{center}
\end{table}

By using the fitted DMH mass distribution and the SMBH mass, we showed the resultant transonic solutions in Figs. \ref{fig-sombrero}a and \ref{fig-sombrero}b. The type of transonic solution is B-1 (Fig. \ref{fig-thisstudy10}). The deduced loci of the critical points of the Sombrero galaxy are shown in Table.~2. Next, we fitted these transonic solutions to the observed gas density distribution within 25\ $\rmn{kpc}$ for the two different temperatures of the gas (0.6 keV and 0.5 keV). In Figs. \ref{fig-sombrero}a and \ref{fig-sombrero}b, the O-point moves inward and X-points move outward when the temperature decreases. This behaviour is the same as that in the analytical results described in Section \ref{result of model}.

We considered the two accelerated transonic solutions of $\rmn{X_{out}}$ and $\rmn{X_{in}}$ as the galactic outflows for fitting the gas density distribution. These accelerated transonic solutions are shown as thick curves in Figs. \ref{fig-sombrero}a and \ref{fig-sombrero}b. By using the results of the fitting, the gas density distributions were determined, as shown in Figs. \ref{fig-sombrero}c and \ref{fig-sombrero}d. In Fig. \ref{fig-sombrero}d, we show that the $\rmn{X_{out}}$ solution in our model can roughly reproduce the observed gas density distribution. For 0.6 keV, the outer transonic point is too far in comparison with the observable region ($<$ 25\ kpc); thus, we were unable to determine the gas density of the $\rmn{X_{out}}$ solution at this temperature. By adding the stellar mass distribution, we can fit the gas density of the $\rmn{X_{out}}$ solution even at 0.6 keV, and the observed gas density distribution can be reproduced at the same temperature, as will be discussed in Section \ref{The Model and Influence of the Stellar Mass Distribution in the Sombrero Galaxy}. The resultant gas density distribution of the type $\rmn{X_{in}}$ solution does not match the observed data and should thus be rejected (Fig. \ref{fig-sombrero}c). This result also holds true for the case with lower temperature (Fig. \ref{fig-sombrero}d). 

As defined in Section \ref{result of model}, the locus of the X-point of type $\rmn{X_{out}}$ is 105.4\ kpc for 0.6 keV and 232.2\  kpc for 0.5 keV. Thus, the subsonic region is very wide in comparison to the observable region ($<$ 25\ kpc).
It should be noted that the gas density distribution in the subsonic region of the type $\rmn{X_{out}}$ solution is quite similar to that of the hydrostatic solution. In addition, the low velocity in the subsonic region was difficult to detect (Fig. \ref{fig-sombrero}d).
Thus, even if observing a physical quantity such as gas density $\rho$ or velocity $v$, we cannot distinguish the type $\rmn{X_{out}}$ solution from the hydrostatic solution.
We can conclude that there is no contradiction between the hydrostatic-like gas density distribution and the widespread hot gas indicating the existence of a galactic outflow \citep{li11}. Thus, we can resolve the problem stated in Section 1.
Our results provide a new perspective of galactic outflow in that slowly accelerating galactic wind may exist even in quiescent galaxies with inactive star formation such as the Sombrero galaxy.

\section{Discussion}

\subsection{Availability of assumptions in our model}
We have assumed that the outflow is spherically symmetric, isothermal and steady without mass injection along the flow lines.
In this section, we discuss the availability of these assumptions.

Mass injection plays roles of mass and energy transport into the flow from stellar winds and supernovae and acts as an effective braking force similar to viscosity.
However, because this injection occurs only in the stellar distribution region, we can neglect it in the widely spread acceleration region of the flow if the transonic point forms outside of the stellar distribution region.
In the Sombrero galaxy, the estimated outer transonic point ($r=127$\ kpc) is far enough from the stellar distribution region ($\leq r_\mathrm{h}=2.53$\ kpc).
Thus, the effect of mass injection is negligible for the Sombrero galaxy in type $\rmn{X_{out}}$ solution.

In X-ray observations, some elliptical galaxies were observed in an isothermal state \citep{fukuzawa06,diehl08}.
Type Ia supernovae were discussed as heat sources in these studies because their feedback has a long time scale compared with that of type II supernovae.
AGN feedback is also considered as a heat source, and thermal conduction is considered as a factor for averaging temperature.
\citet{diehl08} suggested that sufficient thermal conductivity can explain observed temperature structures of elliptical galaxies.
In our model, the $\rmn{X_{out}}$ solution can be accelerated by thermal conduction from the stellar distribution region.
Thus, our main result that the transonic point forms at a distant region may also hold true in polytropic treatment with thermal conduction.

Spherically symmetric approximation is valid for stellar components when $bulge/disk$ mass ratio is large enough because a bulge is spherically symmetric unlike a disk.
In the Sombreo galaxy, the $bulge/disk$ ratio is $\sim 6$ \citep{kent88}; thus, for the gravitational potential, spherically symmetric approximation is relevant for stellar components.

Several studies have predicted the non-spherical distribution for DMH.
For example, \citet{jing02} suggested that a triaxial density model can effectively reproduce the results of {\it N}-body simulations and observed DMH density distribution of dwarf spheroidals.
Although this non-spherical model is still argued by other reseachers, they have not reached the same conclusions as those reached by \citet{jing00,jing02}.
The DMH distributions of dwarf spheroidals predicted by observation indicate widespread ellipticity of $0.3-0.7$ \citep{hayashi12}.
However, we focused on the concentration of DMH that influences the nature of outflows; the asymmetry property is not considered in this study.
Thus, we used the spherically symmetric model for DMH with the variable concentration parameter in our model. 


\subsection{Parameters in actual galaxies}

\subsubsection{$K_\mathrm{DMH}$  \& $K_\mathrm{BH}$}\label{Parameters in Actual Galaxies}
The values of $K_\mathrm{DMH}$, proportional to DMH mass (Eq. (7)), and $K_\mathrm{BH}$, proportional to SMBH mass (Eq. (8)), should be in a plausible range for actual galaxies. Therefore, we estimate the realistic range of these parameters in this section.

First, we estimated the values as follows. The virial temperature $T_\mathrm{v}$ is defined by the relation 
\begin{eqnarray}
&{}&c_\mathrm{s}^2=\frac{k_\mathrm{B}T_\mathrm{v}}{\mu m_\mathrm{H} \eta} \nonumber\\
&{}&\hspace{4mm} =\frac{G(M_\mathrm{DMH}(r_\mathrm{v})+M_\mathrm{BH})}{r_\mathrm{v} \eta} \nonumber\\
&{}&\hspace{4mm} \approx \frac{GM_\mathrm{DMH}(r_\mathrm{v})}{r_\mathrm{v} \eta}, \label{virial temperature}
\end{eqnarray}
where $k_\mathrm{B}$ is the Boltzmann constant, $m_\mathrm{H}$ is the hydrogen mass, $\mu$ is the mean molecular weight, $r_\mathrm{v}$ is the virial radius, $M_\mathrm{DMH}(r_\mathrm{v})$ is the virial mass of DMH (Eq. (\ref{virial mass})) and $\eta$ is the factor for the correction of temperature $T_\mathrm{v}$. In actual galaxies, $\eta$ is believed to be variable at approximately unity \citep{fukuzawa06,diehl08}. Indeed, the virial temperature of the Sombrero galaxy (with $\eta=1$) is $0.9$ keV assuming the virial mass to be $10^{13}\ \mathrm{M}_\mathrm{\odot}$, whereas the observed temperature is $0.6$ keV inside 25\ kpc \citep{li11}. In this case, $\eta$ is $1.5$. At a radius larger than 25\ kpc, the temperature becomes lower than $0.6$ keV \citep{li11}, and $\eta$ becomes larger. 

Next, $M_\mathrm{DMH}(r_\mathrm{v})$ is derived as
\begin{eqnarray}
&{}&M_\mathrm{DMH}(r_\mathrm{v})= 4\pi \int_\mathrm{0}^{r_\mathrm{v}} r^2 \rho_\mathrm{DMH}(r;\alpha,r_\mathrm{d},\rho_\mathrm{d}) dr \nonumber\\
&{}&\hspace{16mm} =4\pi \rho_\mathrm{d} r_\mathrm{d}^3 \int_\mathrm{0}^c x^{2-\alpha}(x+1)^{\alpha-3}dx  \label{virial mass}
\end{eqnarray}
from Eq. (\ref{dmh model}), where $c$ is the concentration parameter defined as $c=r_\mathrm{v}/r_\mathrm{d}$. Several previous studies of {\it N}-body simulations \citep{navarro96,bullock01} concluded the empirical relation between concentration and virial mass of DMH as 
\begin{eqnarray}
c=\kappa  \left(\frac{M_\mathrm{DMH}(r_\mathrm{v})}{10^{12}\ \mathrm{M}_\mathrm{\odot}}\right)^{\xi}, \label{maccio}
\end{eqnarray}
where $\kappa$ and $\xi$ are the different parameters for each study : $(\kappa,\xi)=(12.8,-0.13)\ $\citep{bullock01}, $(9.35,-0.094)\ $\citep{maccio08}, $(9.60,-0.075)\ $\citep{klypin11} and $(9.7,-0.074)\ $\citep{prada12}. It is difficult to observationally determine these parameters; therefore, we adopted the values from \citet{prada12} because their simulation has highest mass resolution.

Furthermore, $M_\mathrm{BH}$ is estimated by using the empirical relation between $M_\mathrm{BH}$ and $M_\mathrm{DMH}(r_\mathrm{v})$ as
\begin{eqnarray}
\left(\frac{M_\mathrm{BH}}{10^8\ \mathrm{M}_\mathrm{\odot}}\right)=\mu \left(\frac{M_\mathrm{DMH}(r_\mathrm{v})}{10^{12}\ \mathrm{M}_\mathrm{\odot}}\right)^{\nu}, \label{ferrarese}
\end{eqnarray}
with $(\mu,\nu)=(0.11,1.27)$ calibrated by \citet{baes03} (see also Ferrarese 2002).
We have adopted the value of \citet{baes03} because they used more observational samples than others. 

Finally, we derive $K_\mathrm{DMH}$ and $K_\mathrm{BH}$ as simple functions of the empirical parameters $\eta$, $c$, $\alpha$, $\nu$ and $\xi$ by the help of Eqs. (\ref{virial temperature})-(\ref{ferrarese}) as
\begin{eqnarray}
&{}&K_\mathrm{DMH}=\eta\frac{c}{2}\left(\int_\mathrm{0}^c x^{2-\alpha}(x+1)^{\alpha-3}dx\right)^{-1}, \label{K and c}\\
&{}&\hspace{2.5mm}K_\mathrm{BH}=\frac{\eta \mu c}{2\times 10^4} \left(\frac{c}{\kappa}\right)^{\frac{\nu-1}{\xi}}. \label{K_{BH} and c}
\end{eqnarray}
We show the actual range of $K_\mathrm{DMH}$ and $K_\mathrm{BH}$ in Figs. \ref{fig-thisstudy10} and \ref{fig-thisstudy00and15}. We determined that the parameters of actual galaxies are in the hatched region of these figures. The four long curves extending horizontally in Figs. \ref{fig-thisstudy10} and \ref{fig-thisstudy00and15} represent the difference in DMH mass. The three curves intersecting these four curves in Figs. \ref{fig-thisstudy10} and \ref{fig-thisstudy00and15} represent the difference in temperature. We determined that the plausible range of the parameters includes the three types, A-1, B-1 and B-2. This result indicates that the locus of galactic outflow is complex such as type B in actual galaxies.

\subsubsection{Ratio of mass fluxes in two types of transonic solutions}

The mass fluxes of the outflows are important to the evolution of galaxies and the release of heavy elements to intergalactic space. 
In this section, we estimate the mass fluxes indicated by our model. If there are two transonic solutions, the mass fluxes are different between two solutions. 
The mass fluxes $\dot{M}$ defined in Eq. (1) are determined by $K_\mathrm{DMH}$, $K_\mathrm{BH}$ and $\rho(x_\mathrm{0})$ as
\begin{eqnarray}
\dot{M}=4\pi v(x_\mathrm{0})\rho(x_\mathrm{0}) r_\mathrm{d}^\mathrm{2} x_\mathrm{0}^\mathrm{2},
\end{eqnarray}
where $v(x_\mathrm{0})$ and $\rho(x_\mathrm{0})$ are the boundary conditions of velocity and gas density distribution at an arbitrary distance $x_0$. 
Because $v(x_\mathrm{0})$ is determined by $K_\mathrm{DMH}$, $K_\mathrm{BH}$ and $x_\mathrm{0}$,
the absolute value of mass flux can be taken arbitrarily in proportion to $\rho(x_\mathrm{0})$.
Thus, we focus on the ratio of the mass fluxes of two transonic solutions with the assumption of the gas distributions of these solutions. We adopt two methods corresponding to this assumption below. 

\begin{figure*}
 \centering
 \includegraphics[width=2\columnwidth]{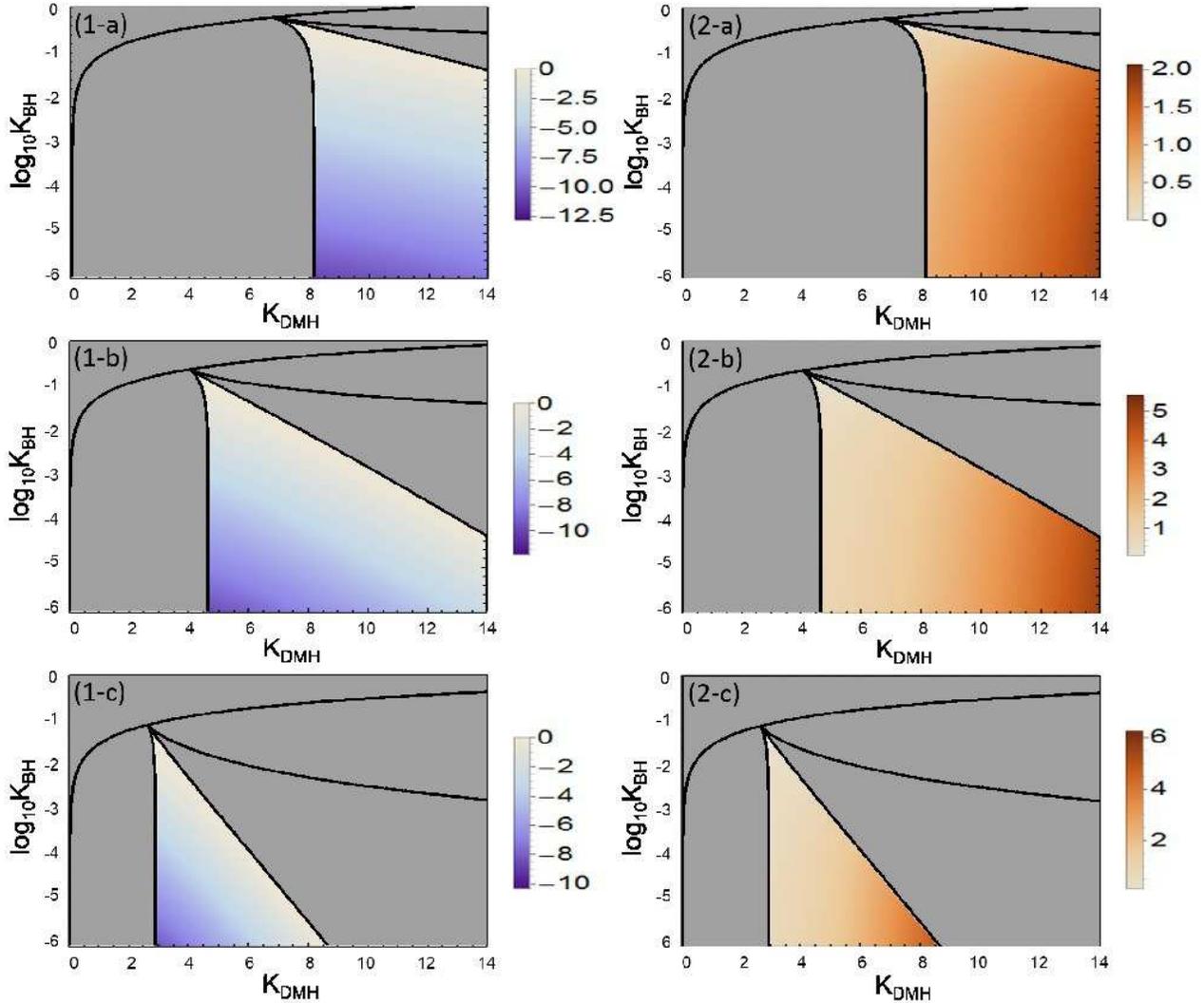}
  \caption{Ratio of mass fluxes in two types of transonic solutions. Colour represents the ratio of mass fluxes $\log_\mathrm{10}\sigma$ defined in Eq. (24). 
The vertical axis is $\log_\mathrm{10}K_\mathrm{BH}$, and the horizontal axis is $K_\mathrm{DMH}$. (1-a): $\alpha=0.0$ with Method 1; (2-a): that with Method 2; (1-b): $\alpha=1.0$ with Method 1; (2-b): that with Method 2; (1-c): $\alpha=1.5$ with Method 1; (2-c): that with Method 2.}
 \label{mass fluxes}
\end{figure*}

\begin{enumerate}
\item[Method (1):] We assume the gas density distributions are hydrostatic-like in the subsonic region. 
The ratio $\sigma$ of mass fluxes reduces
\begin{eqnarray}
&{}&\sigma =\frac{\dot{M}_\mathrm{X_{in}}}{\dot{M}_\mathrm{X_{out}}} \\
&{}&\hspace{2.5mm}=\frac{\rho_{HS}(x_\mathrm{X_{in}})x_\mathrm{X_{in}}^2}{\rho_\mathrm{HS}(x_\mathrm{X_{out}})x_\mathrm{X_{out}}^2} \\
&{}&\hspace{2.5mm}=exp\left\{ \frac{1}{4}(C_\mathrm{X_{out}}-C_\mathrm{X_{in}}) \right\},
\end{eqnarray}
where $x_\mathrm{X_{in}}$ and $x_\mathrm{X_{out}}$ are the loci of the inner X-point and the outer X-point, respectively. 
$\rho_\mathrm{HS}(x_\mathrm{X_{in}})$ and $\rho_\mathrm{HS}(x_\mathrm{X_{out}})$ are the hydrostatic densities at $x_\mathrm{X_{in}}$ and $x_\mathrm{X_{out}}$, respectively. 
$C_\mathrm{X_{out}}$ and $C_\mathrm{X_{in}}$ are the integration constants of the outer transonic solution $\rmn{X_{out}}$ and the inner transonic solution $\rmn{X_{in}}$, respectively. 
These integration constants were determined by Eq. (\ref{thisstudy-mach}) with $\mathcal{M}=1$. 
Therefore, $\sigma$ is determined by three parameters: $\alpha$, $K_\mathrm{DMH}$ and $K_\mathrm{BH}$.

\item[Method (2):] We assume the gas mass in the virial radius is the same in the two transonic solutions. 
The virial radius ($\approx 10r_\mathrm{d}$) is deduced by {\it N}-body simulations (e.g. Navarro et al. 1996). 
The gas mass $M_\mathrm{gas}$ in a galaxy is
\begin{eqnarray}
&{}&\frac{M_\mathrm{gas}}{\mathrm{M}_\mathrm{\odot}}=\zeta\frac{K_\mathrm{DMH}}{K_\mathrm{BH}} \nonumber\\
&{}&\hspace{8.5mm} =\zeta\frac{4\pi\rho_\mathrm{d}r^3_\mathrm{d}}{M_\mathrm{BH}} \nonumber\\
&{}&\hspace{8.5mm} \approx\zeta\frac{M_\mathrm{DMH}}{M_\mathrm{BH}},
\end{eqnarray}
where $\zeta$ is a correction factor. 
In the Sombrero galaxy, $M_\mathrm{gas}\approx 10^8\ \mathrm{M}_\mathrm{\odot}$, $M_\mathrm{DMH}\approx 10^{13}M_{\odot}$ and $M_\mathrm{BH} \approx 10^9\ \mathrm{M}_\mathrm{\odot}$. 
Thus, $\zeta$ is $10^4$ in this galaxy. 
We assume $\zeta=10^4$ in the following discussion. 
In the $\rmn{X_{out}}$ solution of B-1, the gas density distribution of the subsonic region inside the X-point is well described as the hydrostatic state (see Section \ref{the sombrero galaxy}). 
Thus, we assume the flow of $\rmn{X_{out}}$ solution starts from O-point having the gas density estimated by the  hydrostatic state.
\end{enumerate}

As shown by the color contours of $\log_\mathrm{10}\sigma$ in Fig. \ref{mass fluxes}, we have determined that $\sigma$ varies substantially with the three parameters $\alpha,K_\mathrm{DMH}$ and $K_\mathrm{BH}$. 
This result holds true for both methods and indicates that the mass flux is sensitive to variation of these parameters. 
Because of this sensitivity of mass fluxes, the influence of these galactic outflows is hard to estimate, but essential for the evolution of galaxies and the release of heavy elements from them.

\subsection{Review of the Sombrero Galaxy}

\subsubsection{Influence of the stellar mass distribution in the Sombrero galaxy}\label{The Model and Influence of the Stellar Mass Distribution in the Sombrero Galaxy}
In Appendix \ref{The Mass Distribution of DMH in the Sombrero Galaxy}, we estimated the mass distribution of DMH, and we considered DMH and SMBH as the gravity sources in Section \ref{the sombrero galaxy}. However, the stellar component is also an important factor as the gravity source in a galaxy. In this section, we estimate the influence of the stellar mass component.

We adopted the Hernquist model \citep{hernquist90} for the stellar mass distribution, 
\begin{eqnarray}
\rho_\mathrm{H}(r)=\frac{M_\mathrm{H}}{2\pi}\frac{r_\mathrm{h}}{r}\frac{1}{(r+r_\mathrm{h})^2},
\end{eqnarray}
where $r_\mathrm{h}$ is the scale length of the stellar distribution and $M_\mathrm{H}$ is the entire mass of the stellar component. This empirical model effectively reproduces the de Vaucouleurs law \citep{devaucouleurs48}. 

\begin{figure}
 \centering
 \includegraphics[width=0.99\columnwidth]{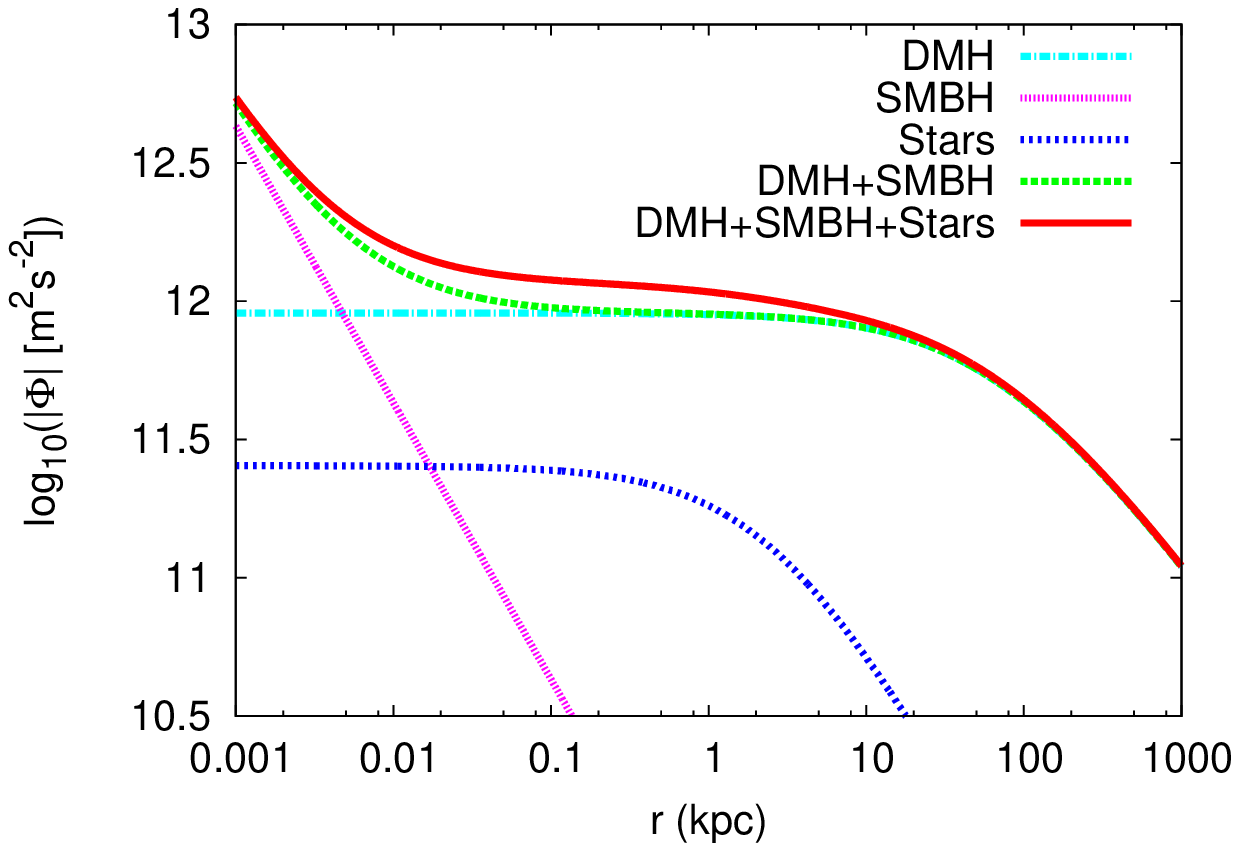}
  \caption{Gravitational potential of the dark matter halo (DMH), super massive black hole (SMBH) and stars. We use parameters of the Sombrero galaxy: $\alpha=1,r_\mathrm{d}=36.1\ \mathrm{kpc}$, $\log_\mathrm{10}M_\mathrm{25}/\mathrm{M}_\mathrm{\odot}= 11.95$, $r_\mathrm{h}=2.53\ \rmn{kpc}$, $M_\mathrm{H}=1.5\times10^{11}\ \mathrm{M}_\mathrm{\odot}$ and $M_\mathrm{BH}=10^9\ \mathrm{M}_\mathrm{\odot}$. The cyan line refers to DMH; the pink line, SMBH; the blue line, stars; the green line, DMH and SMBH; and the magenta line, DMH, SMBH and stars. The gravitational potential of DMH is widespread; that of SMBH is dominant at the central region; and that of stars is concentrated in the inside region. }
 \label{fig-sombreropotential}
\end{figure}

The parameters of the Sombrero galaxy were determined from observation: $r_\mathrm{h}=2.53\ \rmn{kpc}$ and $M_\mathrm{H}=1.5\times10^{11}\ \mathrm{M}_\mathrm{\odot}$ \citep{bell03,bendo06}. In Appendix \ref{The Mass Distribution of DMH in the Sombrero Galaxy}, we estimated the parameters of DMH by fitting to the velocity dispersions of globular clusters \citep{bridges07} with the $\chi^2$ test. Therefore, we adopted $(\alpha,r_\mathrm{d},\log_\mathrm{10}M_\mathrm{25}/\mathrm{M}_\mathrm{\odot})=(1.0,36.1$\ $\rmn{kpc},11.95)$. We show the gravitational potentials of DMH, stars and SMBH in Fig. \ref{fig-sombreropotential}. Because the stellar component is concentrated in the bulge region in approximately $r_\mathrm{h}\approx2.5$\ kpc, the influence of stellar gravity is important in that region.

\begin{figure*}
 \centering
 \includegraphics[width=2\columnwidth]{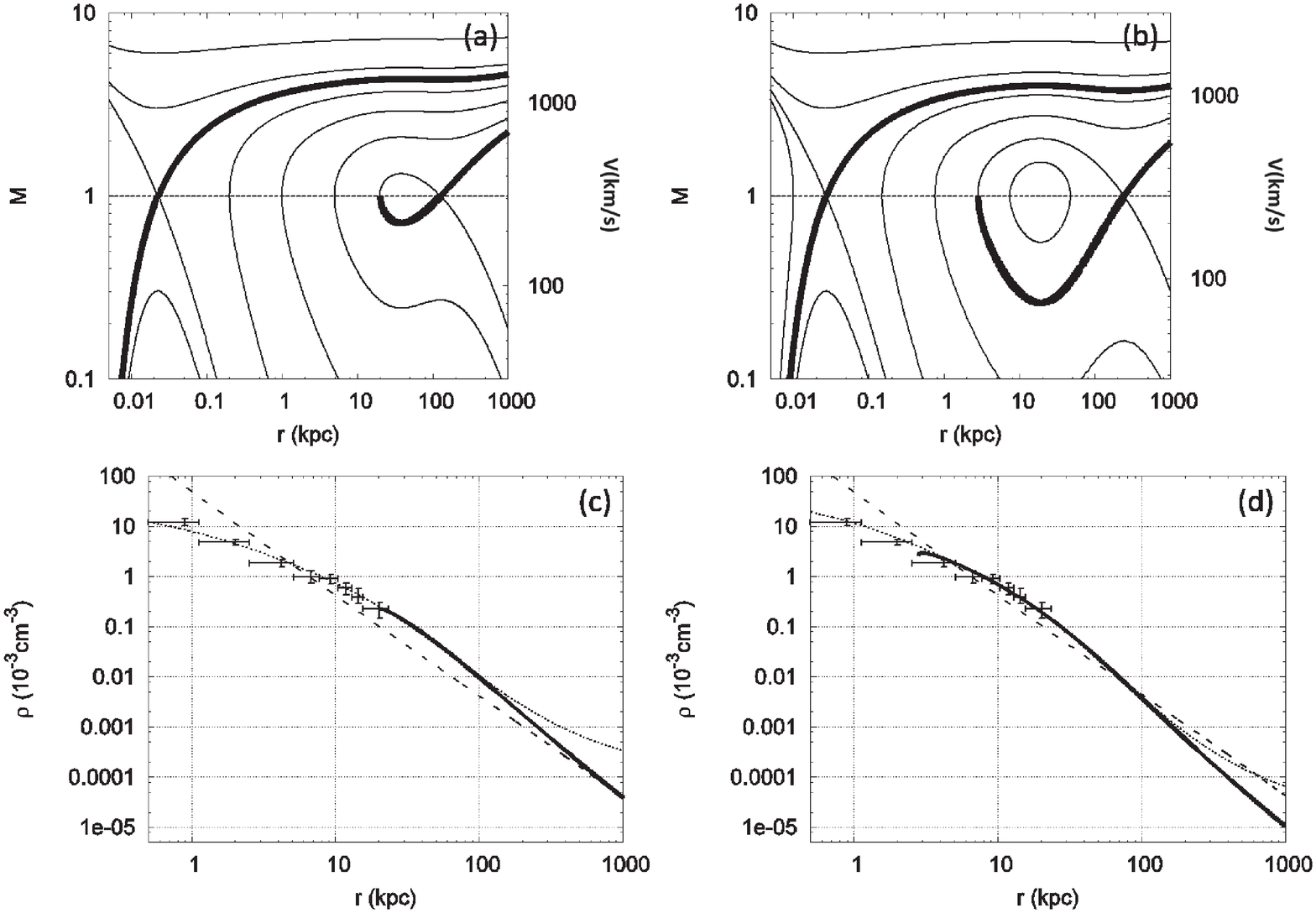}
  \caption{Mach number and gas density distribution in the gravitational potential of the dark matter halo (DMH), super massive black hole (SMBH) and stars. We use parameters of the Sombrero galaxy: $\alpha=1,r_\mathrm{d}=36.1\ \mathrm{kpc}$, $\log_\mathrm{10}M_\mathrm{25}/\mathrm{M}_\mathrm{\odot}= 11.95$, $r_\mathrm{h}=2.53\ \rmn{kpc}$, $M_\mathrm{H}=1.5\times10^{11}\ \mathrm{M}_\mathrm{\odot}$ and $M_\mathrm{BH}=10^9\ \mathrm{M}_\mathrm{\odot}$. The line definitions are the same as those in Fig. \ref{fig-sombrero}.} 
 \label{fig-sombrero2}
\end{figure*}

We show the resultant transonic solutions in Figs. \ref{fig-sombrero2} (a) and (b) by using revised mass distribution including the DMH mass, SMBH mass and stellar mass.
The type of the transonic solution is B-1 (Fig. \ref{fig-thisstudy10}). The revised loci of the critical points of the Sombrero galaxy are shown in Table \ref{tb2}.
The gravitational potential of the stellar component influences the inner region ($\leq$$r_\mathrm{h}\approx2.5$\ kpc), whereas that of DMH is dominant at the far region ($\geq$$r_\mathrm{d}\approx36$\ kpc). Thus, the locus of the outer X-point ($\approx$$200$\ kpc) does not dramatically change in comparison with the case without the stellar component (Tables \ref{tb1} and \ref{tb2}) because the locus of the outer X-point is determined mainly by the gravity of DMH.
Similarly, the locus of that inner X-point ($\approx$$0.02$\ kpc) also does not change in comparison with the case without the stellar component because the gravitational potential of SMBH is dominant in the vicinity of the centre ($\leq$$0.05$\ kpc).

However, the existence of the stellar mass modifies the gravitational potential and moves the O-point inward. Thus, the region covered by the  subsonic region of the $\rmn{X_{out}}$ solution is extended inward. We can determine the $\rmn{X_{out}}$ solutions to the larger number of observed data points ($<$25\ kpc; Li et al. 2011). A comparison of Figs. \ref{fig-sombrero} and \ref{fig-sombrero2} reveals that our model with stellar mass distribution can more effectively reproduce the observed gas density distribution than the model without stellar mass distribution.

The mass flux of the $\rmn{X_{out}}$ solution is $8.77\ \mathrm{M}_\mathrm{\odot}\mathrm{yr}^{-1}$ for $0.6$ $\rmn{keV}$ and $1.84\ \mathrm{M}_\mathrm{\odot}\mathrm{yr}^{-1}$ for $0.5$ $\rmn{keV}$, whereas that of the $\rmn{X_{in}}$ solution is $17.0\ \mathrm{M}_\mathrm{\odot}\mathrm{yr}^{-1}$ for $0.6$ $\rmn{keV}$ and $14.6\ \mathrm{M}_\mathrm{\odot}\mathrm{yr}^{-1}$ for $0.5$ $\rmn{keV}$.
These values are much larger than the estimated mass injection of approximately $0.3\ \mathrm{M}_\mathrm{\odot}\mathrm{yr}^{-1}$ from the stellar component in the Sombrero galaxy \citep{athey02,knapp92}.
This discrepancy implies several new aspects for outflows.
For example, if the opening angle of the outflow is smaller than $\pi$ (i.e., collimated bipolar outflow), the mass flux is smaller than that of the spherical outflow.
Indeed, the outflow that has been observed as a bipolar-like structure vertical to the disk \citep{li11}.
If outflows occur intermittently, it is not a problem that the mass flux is larger than the supply.
If an outflow is intermittent, there are active periods for galactic outflows and inactive periods for gas storage in galaxies.
When the amount of gas reaches a certain value, an outflow occurs.

\begin{table}
 \centering
 \caption{The critical points with the gravitational potential of DMH, SMBH and stars}\label{tb2}
  \begin{tabular}{|c|c|c|c|}
    &$r_\mathrm{BH}(\rmn{kpc})$&$r_\mathrm{O}(\rmn{kpc})$&$r_\mathrm{DMH}(\rmn{kpc})$ \\ \hline
   $0.6\rmn{keV}$&0.021&38.2&126.9\\ \hline
   $0.5\rmn{keV}$&0.026&18.8&243.2\\ \hline
  \end{tabular}
\end{table}

\subsubsection{Deduction of mass profile from outflow velocities}
The results given in Section \ref{Parameters in Actual Galaxies} show that the mass profile strongly affects the acceleration process of outflows.
Thus, we can deduce the mass distribution from the velocity distribution of galactic outflow if it can be observed.
For example, for type B-1 in Fig. 2, the flow of the $\rmn{X_{in}}$ solution is accelerated in the central region under SMBH gravity, whereas the flow of the $\rmn{X_{out}}$ solution is accelerated in the distant region under DMH gravity.
Thus, we can deduce the mass of SMBH from observed velocity distribution for the case of $\rmn{X_{in}}$ solution, similarly we can deduce the mass distribution of DMH from observed velocity distribution for the case of $\rmn{X_{out}}$ solution.
In the current technology of X-ray observation, it may be difficult to detect very slow outflow ($\simeq$ 100 km s$^{-1}$) of hot gas in detail, although future missions such as {\it ASTRO-H} may enable us to detect it.
Beyond the traditional analysis based on the kinematics, our method provides a new technique to estimate the mass distributions of DMH and SMBH.

\section{Conclusion}
We have revealed possible transonic solutions in the gravitational potential of DMH and SMBH by using an isothermal, spherically symmetric and steady model. These solutions are classified from the perspective of their topological features in the phase diagram shown in Fig. 2.
We conclude that gravitational potential of SMBH adds a new branch of transonic solution with the inner transonic point compared with the transonic point generated by the gravity of DMH, whereas the previous research \citet{tsuchiya13} concluded that there is only one transonic solution in the gravitational field of DMH.
We have analysed the relation between the properties of each gravity source generating transonic solutions.
We conclude that the inner transonic point is generated by the gravity of SMBH, whereas the outer transonic point is generated by the gravity of DMH.

The most interesting new feature of this study is the suggestion of two different types of transonic solutions with substantially different mass fluxes and starting points (Figs. 2 and \ref{mass fluxes}).
Thus, these solutions may have different influences on the evolution of galaxies and the release of metals from them.
In addition, we have estimated the parameter ranges of actual galaxies by using the results of previous studies.
We have determined that actual galaxies can have two transonic points.

We have applied our model to the Sombrero galaxy and have shown the possibility of galactic outflow in this galaxy.
The distribution of DMH is estimated from the observation of globular cluster kinematics.
The transonic solution through the outer transonic point ($\sim126$\ kpc) has a quite similar gas density profile as that of the hydrostatic solution in the widespread subsonic region.
Therefore, the existence of the slowly accelerating outflow may support the result of the previous work reporting the coexistence of hydrostatic-like gas distribution and widespread hot gas \citep{li11}.
Because the Sombrero galaxy is not an active star-forming galaxy, this possibility of slowly accelerating outflow indicates that galactic outflow could exist for other quiescent galaxies without active star formation.
Although the majority of theoretical researches of galactic outflows have focused on the star-forming galaxies, our results provide new perspectives of galactic outflows that are applicable to quiescent galaxies without active star formation. Indeed, observational studies reported the possible outflows from quiescent galaxies including M31 \citep{li07_2,bogdan08} and NGC4278 \citep{pellegrini12,bogdan12}. We will apply our model to these objects in the series of our forthcoming studies.

We can estimate the mass distributions of DMH and SMBH from the structure of outflow velocity distribution by using our model.
The transonic solution through the outer transonic point suggests mass distribution of DMH, whereas the transonic solution through the inner transonic point suggests mass of SMBH.
Thus, we can estimate the mass distribution of a galaxy from the observation of outflow velocity.
Previous studies (e.g. Bridges et al. 2007) estimated mass distribution by using stellar kinematics or orbital motions of globular clusters.
However, the method for the mass distribution estimation presented in this study differs from that in previous methods.
Although the velocity of hot gas is difficult to estimate through current X-ray observation techniques, the next generation of X-ray observation will be able to detect detailed structures of galactic hot gas including velocity.
We will attempt to estimate the mass distribution from galactic outflow velocity with a more realistic model in future work. 

\section*{Acknowledgements}
This work was supported in part by JSPS Grants-in-Aid for Scientific Research: (A) (21244013) and (C) (25400222, 20540242). For Fig. 4, we used S2PLOT software \citep{barnes08,fluke08}.

\appendix
\section{Mass distribution of DMH in the Sombrero galaxy} \label{The Mass Distribution of DMH in the Sombrero Galaxy}
For the application of our model to the Sombrero galaxy, the mass distribution of this galaxy must be determined. The main components of mass distribution are DMH, SMBH and stars. The SMBH mass of the Sombrero galaxy is estimated as $10^9\ \mathrm{M}_\mathrm{\odot}$ \citep{kormendy96}. This SMBH is believed to be inactive \citep{heckman80}. 

We estimated the parameters of DMH by fitting to the velocity dispersions of globular clusters \citep{bridges07} by using the $\chi^2$ test. Thus, we determined the values of the three parameters ($\alpha,r_\mathrm{d},\rho_\mathrm{d}$) in Eq. (\ref{dmh model}). In place of the scale density $\rho_\mathrm{d}$, we used the total mass $M_\mathrm{25}$ of DMH inside 25\ kpc, which was determined as
\begin{equation}
M_\mathrm{25}= 4\pi \int_\mathrm{0\ \rmn{kpc}}^{25\ \rmn{kpc}} r^2 \rho_\mathrm{DMH}(r;\alpha,r_\mathrm{d},\rho_\mathrm{d}) dr 
\end{equation}
because the total mass is more easily comprehensible than the scale density. 

Because we fixed the DMH mass and the SMBH mass as the sources of the gravitational potential in our model, we deduced the stellar mass from the observed mass distribution by using the observed stellar mass distribution, as discussed in Section \ref{The Model and Influence of the Stellar Mass Distribution in the Sombrero Galaxy}. Thus, we determined the pure DMH mass distribution excluding the stellar mass and SMBH mass. 

\citet{li11} also calculated the mass distribution of DMH by using the same observed data \citep{bridges07}, although we estimated our data by a different method. First, they used the NFW model. Thus, from the beginning, they were not able to consider the concentration of DMH. On the other hand, our model of DMH distribution in Eq. (\ref{dmh model}) can reproduce the various mass distributions by varying $\alpha$ (Section 2.1). Second, although the observed data represents the cumulative mass distribution, the value decreases in the far distance. Thus, we used the data inside the peak ($\approx$25\ kpc), whereas \citet{li11} used the data of the entire region. Third, they neglected the mass of SMBH; we added this gravitational potential of SMBH to our model. 

\begin{figure}
\begin{center}
 \includegraphics[width=0.99\columnwidth]{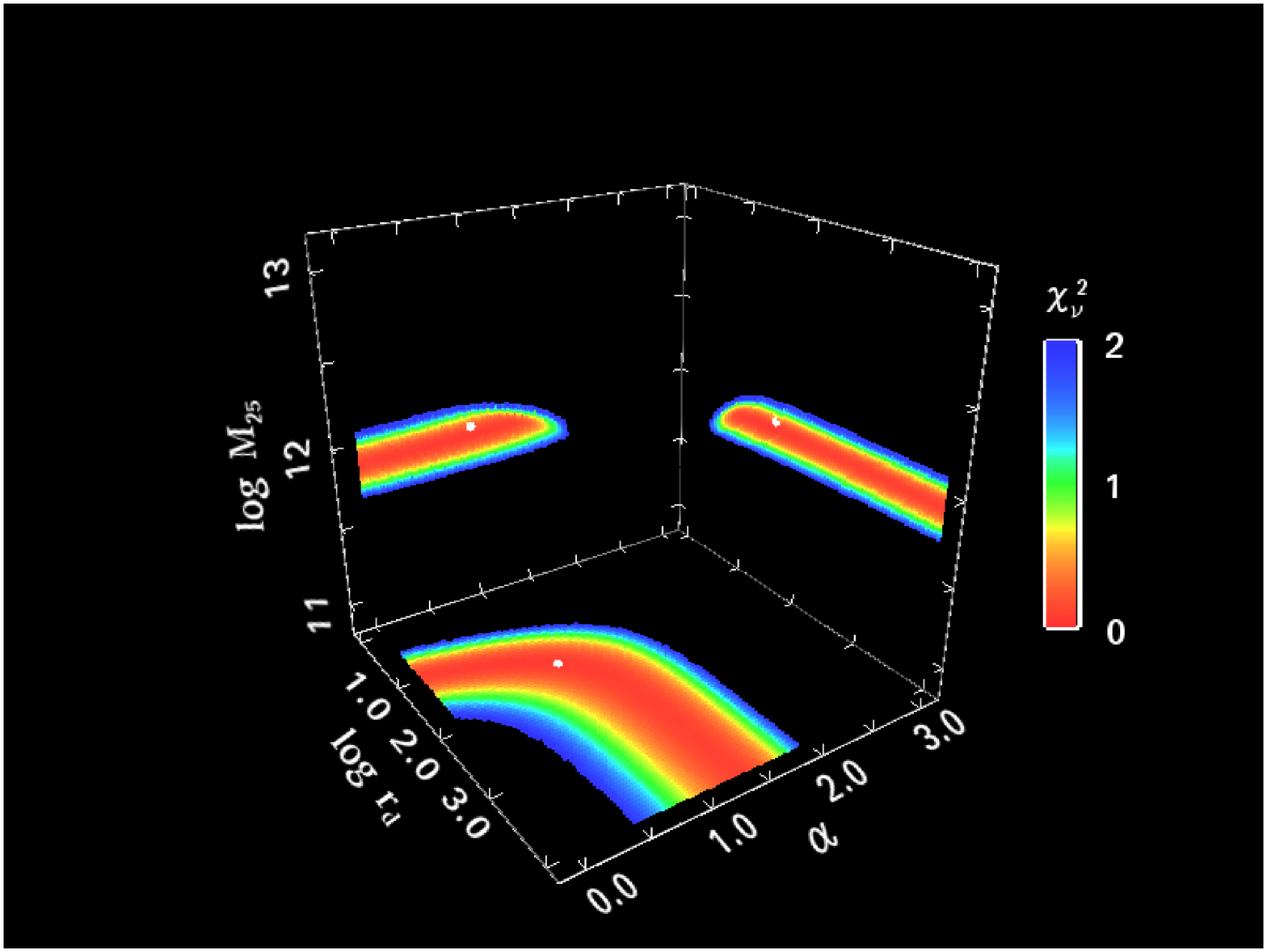}
  \caption{$\chi_{\mathrm{\nu}}^2$ map of fitting for ($\alpha,r_\mathrm{d},M_\mathrm{25}$). The vertical axis shows $\log_\mathrm{10}(M_\mathrm{25}/\mathrm{M}_\mathrm{\odot})$ and the horizontal axes show $\alpha$ and $\log_\mathrm{10}r_\mathrm{d}(kpc)$. 
White dots indicate the adopted parameters in this study. } 
 \label{fig-fitting}
\end{center}
\end{figure}

Fig. \ref{fig-fitting} shows the $\chi_{\mathrm{\nu}}^2$ map as a function of ($\alpha,r_\mathrm{d},M_\mathrm{25}$). $\chi_{\mathrm{\nu}}^2$ is reduced to a value less than unity in the wide region from $\alpha=0$ to $\alpha\approx 1.5$, but increases rapidly in $\alpha \geq 2$. The minimum $\chi^2$ is in the vicinity of $\alpha=1.0$. 
In this paper, we adopted $(\alpha,r_\mathrm{d}/ \rmn{kpc},\log_\mathrm{10}M_\mathrm{25}/\mathrm{M}_\mathrm{\odot})=(1.0,36.1,11.95)$, and $\chi_{\mathrm{\nu}}^2=0.06$ in this case.
This distribution is more widely spread and more massive than that reported by \citet{li11}. 

\bsp

\label{lastpage}

\end{document}